\documentclass[a4paper,10pt]{article}
\usepackage[utf8]{inputenc}
\usepackage{amssymb}
\usepackage{amsmath}
\usepackage[a4paper,left=25mm,right=25mm,top=25mm,bottom=25mm]{}
\usepackage[english]{babel}
\usepackage[nottoc]{tocbibind}
\usepackage{biblatex}
\usepackage{hyperref}
\newcommand*{\QEDB}{\null\nobreak\hfill\ensuremath{\square}}%
\title{\normalsize{\textbf{A NEW QUANTUM ALGORITHM FOR THE HIDDEN SHIFT PROBLEM IN $\mathbb{Z}_{2^t}^n$}}}
\author{\footnotesize{GERGELY CSÁJI}\\
\footnotesize{\textit{Eötvös Loránd University, 1117 Budapest, Pázmány Péter sétány 1/A}}\\
\footnotesize{\textit{Budapest, 1185, Hungary}}\\
\footnotesize{\textit{csaji.gergely@gmail.com}}
}
\usepackage{xcolor}

\begin{document}
\maketitle
\begin{abstract}
\baselineskip=10pt
   \footnotesize{ In this paper we make a step towards a time and space efficient algorithm for the hidden shift problem for groups of the form $\mathbb{Z}_k^n$. We give a solution to the case when $k$ is a power of 2, which has polynomial running time in $n$, and only uses quadratic classical, and linear quantum space in $n\log (k)$. It can be a useful tool in the general case of the hidden shift and hidden subgroup problems too, since one of the main algorithms made to solve them can use this algorithm as a subroutine in its recursive steps, making it more efficient in some instances.}
\end{abstract}
\section{Introduction}
\textwidth=5in
\textheight=7.8in
\baselineskip=13pt
The hidden subgroup and hidden shift problems 
have been intensively studied by several authors since
Shor's discovery of efficient factoring and discrete logarithm
algorithms\textsuperscript{\cite{DBLP:journals/siamcomp/Shor97}}. 
Many
of the problems that have an exponentially faster quantum algorithm are instances of the first one and the latter is a closely related problem, which is useful for example when we are dealing with the hidden subgroup problem in groups of the form $G\rtimes \mathbb{Z}_2$, $G$ abelian. \\
The hidden subgroup problem consists of a finite group $G$, a subgroup $H\le G$, a finite set $S\subset\{0,1\}^l$, (this $l$ is called the encoding length) and a function $f:G\to S$ such that $f(x)=f(y)$ if and only if $x$ and $y$ are both elements of the same left coset of $H$. We then say that $f$ hides the subgroup $H$. Furthermore we suppose that $f$ is efficiently computable or that we have an oracle $U_f$ mapping the state $|x\rangle |0\rangle$ to $|x\rangle |f(x)\rangle$. 
\\
In the hidden shift problem we are given a finite group $G$, a finite set $S\subset\{0,1\}^l$, and two functions $f_0,f_1:G\to S$, both injective and we know that there exists some $s\in G$ such that $f_0(x)=f_1(xs)$ for every $x\in G$. The task is to find $s$, given oracle access to
$f_0$ and $f_1$, which are unitary transforms $U_{f_0},U_{f_1}$, that map the state $|x\rangle |0\rangle$ to $|x\rangle |f_i(x)\rangle$. 
\\
For example, if we have a finite abelian group $G$, then the hidden shift problem can be viewed as an instance of the hidden subgroup problem with $G'=G\rtimes \mathbb{Z}_2$, and $f:G'\to S$ given with $f(x,i)=f_i(x)$. Then the function $f$ hides the subgroup $H=\{ (0,0), (s,1)\}$, so if we find $H$, we can determine $s$ too. Ettinger and Hoyer showed\textsuperscript{\cite{DBLP:journals/aam/EttingerH00}} that there is a reduction in the other direction as well, therefore the two problems are quantum polynomially equivalent.
 \\ There has been a massive amount of research regarding these problems, however an efficient solution to the general case has not yet been discovered and
is starting to seem impossible. But in some special cases we can at least give some algorithms, that are quite more efficient than the brute force search, which is generally the only classical solution. For example when $G$ is abelian, then we can solve the hidden subgroup problem in polynomial time, but the hidden shift problem seems more difficult even in this case.
\\
The hidden shift problem, although not as general as the hidden subgroup problem, is still very useful in a lot of areas. 
For example, as Regev showed\textsuperscript{\cite{doi:10.1137/S0097539703440678}}, the case when $G=\mathbb{Z}_n$ (which corresponds to the hidden subgroup problem in the dihedral group) is important, 
because an efficient
fault tolerant
algorithm for this would 
break a primitive from lattice based cryptography. 
Another relevant aspect for the hidden shift problem is that 
it is
recursively used by an algorithm of Friedl et al.\textsuperscript{\cite{DBLP:journals/siamcomp/FriedlIMSS14}} which is one of the most important algorithms for the hidden subgroup problem (it can be used to solve the hidden shift problem as well), because it works efficiently for 
a reasonably large
class of groups, namely solvable groups with constant exponent 
and
constant derived 
length. It introduces a new problem called \textit{Translating Coset}, which is a generalization of both the hidden subgroup and the hidden shift problems and then the authors show that this \textit{Translating Coset} in any finite solvable group $G$ and for any normal subgroup $N$ of $G$ is reducible to the \textit{Translating Coset} in $G/N$ and in $N$. The hidden shift problem comes into the picture, because the algorithm used to solve \textit{Translating Coset} uses an algorithm for the hidden shift problem in $\mathbb{Z}_p^n$ as a subroutine. So more efficient algorithms for the hidden shift problem in these cases or efficient algorithms for a larger class of groups, for example $\mathbb{Z}_{p^t}^n$ would mean a more efficient solution in the general case too, the latter because then the algorithm 
could save $t-1$ 
recursive calls. Our paper adresses this problem and gives an efficient solution to the case when $G=\mathbb{Z}_{2^t}^n.$
\\
Generally there are two main algorithms for the hidden shift problem. The first one is Kuperberg's\textsuperscript{\cite{DBLP:journals/siamcomp/Kuperberg05}}, which solves the problem in $2^{\mathcal{O}(\sqrt{\log |G|})}$ time and space. 
It is currently considered the best 
algorithm for the 
general case of the
problem. 
Although Kuperberg's algorithm used exponential space, Regev made a modified version, which uses a pipeline of routines, each waiting until it gets enough objects from the previous one and reduces the space complexity to only polynomial in $\log |G|$.
\\ The other is the previously mentioned FIMMS algorithm of Friedl et al.\textsuperscript{\cite{DBLP:journals/siamcomp/FriedlIMSS14}} Its running time is 
$((\log |G|+e+2^{\sqrt{\log (s)}})^{\mathcal{O} (e)}\log (1/\varepsilon ))^r$ 
(see Ref.~\cite{DBLP:journals/siamcomp/FriedlIMSS14} Theorem 4.13), if $G$ has a subnormal series of length $r$, where each factor is either elementary abelian of prime exponent bounded by $e$ or is an abelian group of order at most $s$ and $\varepsilon$ is the allowed error probability. 
So it works efficiently in  solvable groups with constant exponent 
and
constant derived
lenght. 
\\ The space complexity is exponential, because the algorithm has to have all the states it needs at the beginning, because making further copies would violate the no cloning theorem. 
\\
In this paper we give a new algorithm, that solves the hidden shift problem in $\mathbb{Z}_{2^t}^n$. It combines some ideas from previous algorithms to achieve both polynomial space complexity and polynomial running time in $n$. So it is faster than Kuperberg's algorithm in some cases, for example when $t$ is constant, or is asymptotically small compared to $n$. The FIMSS algorithm has similar running time, but its exponential space complexity is a lot worse than our quadratic space need.
So ours is the first algorithm for the hidden shift problem in groups of the form $\mathbb{Z}_k^n$, that has both polynomial running time in $n$ and polynomial space complexity (apart from Simon's algorithm for the case $G=\mathbb{Z}_2^n)$. We hope its ideas can be extended to make more general algorithms with similar efficiency. 
\\ Oddly enough, space complexity hasn't had too much attention in the literature of quantum computing, despite the fact that even if we can build quantum computers, that can work like the mathematical model we use to describe them, they probably won't have too much space to operate with, especially in the beginning, because the more qubits we have, the more difficult to maintain the quantum states of qubits and prevent errors, decoherence and unwanted entaglement, so it's important to make algorithms, that will be able to run with these limitations. \\
Now we state our main result:
\\ \\
\textbf{Theorem 1.} \textit{There is a quantum algorithm, that solves the hidden shift problem in $\mathbb{Z}_{2^t}^n$ with success probability $1-\varepsilon$ (for any $\varepsilon >0)$ in $\mathcal{O}(t^3(n+1)^{(t+2)}l)\log (\varepsilon^{-1} )$ time and $\mathcal{O}(tn^2+l)$ classical and $\mathcal{O}(tn+l)$ quantum space.}
\\ \\
The structure of this paper is the following. In Section 2 we describe the new algorithm in details.
In Section 3 we analyse the algorithm as well as prove its correctness and clarify some of the methods mentioned in Section 2. In the end of Section 3 we give a modified version of our algorithm that runs with coset states as input.
\\It turns out that the FIMSS algorithm can be tailored for the special
case $\mathbb{Z}_{2^t}^n$ to achieve very close running time to our method's, 
although 
the space complexity remains exponential in $t$. We outline
such a version of the FIMSS in Section 4.

\section{The new algorithm}

In this section we present the algorithm for solving the hidden shift problem in $\mathbb{Z}_{2^t}^n$ or equivalently the hidden subgroup problem in $\mathbb{Z}_{2^t}^n\rtimes \mathbb{Z}_2$, with hidden subgroup $H=\{ (0,0),(s,1)\} $ and $f:\mathbb{Z}_{2^t}^n\rtimes \mathbb{Z}_2 \to S$, $f(x,i)=f_i(x)$. It has some of its main ideas based on Kuperberg's subexponential-time algorithm\textsuperscript{\cite{DBLP:journals/siamcomp/Kuperberg05}} for solving the dihedral hidden subgroup problem in $\mathbb{Z}_{2^t}\rtimes \mathbb{Z}_{2}$. Although it becomes exponential in $t$, it is only polynomial in $n$, so in the case when $t$ is constant or $n$ is asymptotically large compared to $t$, it
is 
more efficient than Kuperberg's general algorithm for this problem. \\ \\\\
\textbf{Algorithm:} \\
\textbf{Input:} $N,n\in \mathbb{Z}$, where $N=2^t$.\\
\textbf{Oracle input:} $f:\mathbb{Z}_{2^t}^n\rtimes \mathbb{Z}_2 \to S\subset\{0,1\}^l$, hiding $H=\{ (0,0),(s,1)\} $ for some $s\in \mathbb{Z}_{2^t}^n$.
\\
\textbf{Output:} $s$.
\\ \\
First, we prepare a state in $|0^{(tn+1+l)}\rangle$. Then we apply Hadamard operations to the first $tn+1$ qubits to achieve the uniform superposition state:
$$\frac{1}{\sqrt{2^{tn+1}}}\sum_{x \in \mathbb{Z}_{2^t}^n} (|x\rangle |0\rangle  + |x\rangle |1\rangle )|0^l\rangle. $$
We then apply the unitary transform $U_f$ 
(which maps the state $|Z\rangle |0\rangle$ to $|Z\rangle |f(Z)\rangle$) 
to get:
$$\frac{1}{\sqrt{2^{tn+1}}}\sum_{x \in \mathbb{Z}_{2^t}^n} (|x\rangle |0\rangle |f(x,0)\rangle + |x\rangle |1\rangle |f(x,1)\rangle ).$$
We then measure the last $l$ bits and assuming we measured $f(x,0)$ for some $x\in \mathbb{Z}_{2^t}^n$, then omitting the last $l$ bits the state becomes:
$$\frac{1}{\sqrt{2}}(|x\rangle |0\rangle + |x+s\rangle |1\rangle ).$$
Next we use a quantum Fourier-transform (QFT) over $\mathbb{Z}_{2^t}^n$ to obtain the state:
$$\frac{1}{\sqrt{2^{tn+1}}}\sum_{u\in \mathbb{Z}_{2^t}^n} (e^{\frac{2\pi i\langle u,x\rangle}{N}}|u\rangle |0\rangle + e^{\frac{2\pi i\langle u,x+s\rangle}{N}}|u\rangle |1\rangle ),$$
which can be written as:
$$\frac{1}{\sqrt{2^{tn+1}}}\sum_{u\in \mathbb{Z}_{2^t}^n}e^{\frac{2\pi i\langle u,x\rangle}{N}}|u\rangle \otimes (|0\rangle +e^{\frac{2\pi i\langle u,s\rangle}{N}}|1\rangle ).$$
Then, after a measurement on the first $tn$ qubits, we will obtain a state similar to the ones in Kuperberg's algorithm:
$$|\phi_u \rangle =\frac{1}{\sqrt{2}}|0\rangle +e^{\frac{2\pi i\langle u,s\rangle}{N}}|1\rangle ,$$
with the value of $u$ being known. Then we generate $n+1$ 
of them the same way.
The rest of the algorithm will go as follows:
\\ \\
\textit{Step 1} Having
$(n+1)$ of the $u$-s, we can find coefficients $a_1$,...,$a_{n+1}\in \{0,1\}$, such that $\sum_{i=1}^{n+1} a_iu_i\equiv 0 \;(mod\; 2)$, because any $n+1$ vectors have to be linearly dependent
in 
$\mathbb{Z}_2^n$. If there are more than one option for the choice of $\textbf{a}=(a_1,...,a_{n+1})$, then we will select one randomly with equal probabilities (the method will be clarified in the next section).
\\ \\
\textit{Step 2}  Using Kuperberg's method (see Section 4), from two states $|\phi_u\rangle$ and $|\phi_{u'}\rangle$ with a suitable measurement we can extract a state $|\phi_{u+u'}\rangle$ or $|\phi_{u-u'}\rangle$ with equal probabilities. With that technique we will "add" the $|\phi_u\rangle$ states corresponding to those $a_i$-s, that were equal to $1$ to get $|\phi_{v^{(1)}} \rangle$ states, where $v^{(1)}$ has coordinates divisible by 2 and is uniformly distributed over $2\mathbb{Z}_{2^t}^n=\mathbb{Z}_{2^{t-1}}^n$ (as we will show later). Since $u+u'$ and $u-u'$ have the same parity, obtaining $v^{(1)}=\sum_{i=1}^{n+1} \varepsilon_iu_i$, $\varepsilon_i\in \{-1,1\}$ instead won't be a problem, as it will be congurent to $0$ and the uniformity will remain
too,
as we will see.
\\ \\
\textit{Step 3} We divide each of the achieved sums by 2 to obtain an other sample of $u^{(1)}=\frac{v^{(1)}}{2}$-s. \\
When we are out of $u$-s we will generate another $n+1$, until we have $n+1$ of $u^{(1)}$-s and $|\phi_{v^{(1)}}\rangle$-s too.
\\ \\
 We repeat \textit{Step 1}, \textit{Step 2} and \textit{Step 3} with the new sample of $u^{(1)}$-s, then with $u^{(2)}$-s, and so forth, until we have a sample of $(n+t)$ $u^{(t-1)}$ elements, uniformly distributed over $\mathbb{Z}_2^n$.
\\
Notice that we use a pipeline like routine similar to Regev's\textsuperscript{\cite{article}},  so at each level we wait until we have enough states for the next step, so this way we won't need more than 
$n(t-1)+n+t=nt+t$
states at the same time ever during the algorithm.
\\
\\
\textit{Step 4}  The final $|\phi_{v^{(t-1)}}\rangle$ states are of the form $\frac{1}{\sqrt{2}}(|0\rangle + (-1)^{\sum_{i\in \mathcal{I}}s_i}|1\rangle )$, where $\mathcal{I}=\{ i\in \{1,...,n+1\} :v^{(t-1)}_i=2^{t-1}\}$, because if we add the original $u$-s contained in each $u^{(t-1)}$, then the sums will consist of elements, that have each of their coordinates either $0$ or $2^{t-1}$. So now a measurement in the $|\pm \rangle$ basis (the one with basis vectors $\frac{1}{\sqrt{2}}(|0\rangle +|1\rangle )$ and $\frac{1}{\sqrt{2}}(|0\rangle -|1\rangle )$) reveals the parity of $\sum_{i\in \mathcal{I}}s_i$. And with $(n+t)$ states, we will have $n$ linearly independent equations with probability very close to 1 (This probability is at least $ (1-\frac{1}{2^{t}})$, since its $1-Pr$(any $n$ are linearly dependent)$\ge 1-2^nPr($ every point are from a given $(n-1)$-dimensional subspace of $\mathbb{Z}_2^n$)$=1-2^n(\frac{1}{2})^{n+t}=1-\frac{1}{2^t}$), so we can determine the vector $s$ $(mod$ $2$).
\\ \\
\textit{Step 5} Knowing $s$ $(mod$ $2$) we can pass on to a subgroup $G_1$ of $\mathbb{Z}_{2^t}^n\rtimes \mathbb{Z}_2^n$ isomorphic to $\mathbb{Z}_{2^{t-1}}^n\rtimes \mathbb{Z}_2^n$ containing each possible hidden subgroup $H$ (we will give the detailed method in the next chapter) and repeat the algorithm from \textit{step 1} for $G_1$ to obtain $s$ $(mod$ $4)$, then for $G_2$,...,$G_{t-1}$ to get the exact value of $s$. The probability that we can determine $s \; (mod$ $2^i)$, for every $i=1,...,t$ is at least $(1-2^{-t})^t\ge \frac{1}{2}$. So $\mathcal{O}(\log (\varepsilon^{-1} ))$ applications is enough to obtain $s$ with $1-\varepsilon$ probability
\\
\\
In order to achieve the linear quantum space usage of our algorithm, notice that the $|\varphi_u\rangle$ states only require one qubit and it's the $u$ vectors and the calculations with them that require the most space. But since the $u$ vectors can be described by classical bits and the calculation with them can be done on a classical computer we can just do this part of our algorithm on a classical computer and only use the quantum computer to create, add and measure the $|\varphi_u\rangle$-s. We can even store the $u$-s and $|\varphi_u\rangle$-s in an array such that the $i$-th cell of the array of the classical machine contains $u_i$ and the $i$-th cell of the quantum computer's array contains $|\varphi_{u_i}\rangle$, so we can trace back what $|\varphi_u\rangle$-s to add.
Therefore after we created a $(u,|\varphi_u\rangle )$ pair in $\mathcal{O}(nt+l)$ space, we can copy this classical $u$ to a classical computer and use the same storage for creating the next pair only excluding the one qubit that $|\varphi_u\rangle$ uses. So generating $\mathcal{O}(n)$ pairs can be done with $\mathcal{O}(nt+l)$ quantum space and so does the whole algorithm.
\section{Analysis of the algorithm}
\subsection{The methods and theorems needed}

In this section we will give details of the methods used in our algorithm, as well as prove some lemmas needed for its correctness.
\\ 
First we will prove, that in \textit{step 2} we get a uniformly random element from $\mathbb{Z}_{2^{t-1}}^n$ indeed.
\\ \\
\textbf{Lemma 1.}\textit{ Let $u_1,...,u_{n+1}$ be a uniformly random sample from $\mathbb{Z}_{2^t}^n$, and let $A=\{\textbf{a}=(a_1,...,a_{n+1})\ne (0,...,0) \; :\; \sum_{i=1}^{n+1}a_iu_i\equiv 0 (mod \; 2), \; a_i\in \{0,\pm 1\} \}  $. Furthermore, choose an $\textbf{a}\in A$ randomly (each with equal probabilities). Then $\frac{\sum_{i=1}^{n+1}a_iu_i}{2}$ is an uniformly random element from $\mathbb{Z}_{2^{t-1}}^n$.}
\\ \\
\textbf{Proof} Notice that for any $\emptyset \ne \mathcal{I}\subset \{1,...,n+1\}$ the sum $\sum_{i\in \mathcal{I}}\varepsilon_iu_i$, $\varepsilon_i\in \{-1,1\}$ is a uniformly random element from $\mathbb{Z}_{2^t}^n$, because adding or subtracting the last $u_i$ can yield any vector with equal probabilities, since $u_i$ is uniformly random. 
\\
Next, let $x$ be an arbitrary element in $\mathbb{Z}_{2^{t-1}}^n$. Then
$$P(\frac{\sum a_iu_i}{2}=x)=P(\sum a_iu_i=2x)=P((A_1\cap B_1)\cup (A_2\cap B_2)\cup ...\cup (A_m\cap B_m)),$$
where $A_i$ is the event that we choose the $i$-th element of $A$ (after ordering $A$'s elements some way) and $B_i$ is the event $\{ \sum_{j\in \mathcal{J}}a_ju_j=2x\}$, where $\mathcal{J}$ consists of those $j$-s, for which $a_j=\pm 1$ in the $i$-th element of $A$.
\\
Now observe, that the probability of $B_i$ doesn't depend on the choice of $x$ and that fixing $u_i$-s for some $\mathcal{J}\subset \{ 1,..,n+1\}$, the rest of the $u_i$-s will be independent of that, and because the probability that we choose a given vector $\mathbf{a}$ only depends on the parity of the coordinates of the $u_i$-s (since it only depends on what elements will $A$ consist of) and they have the same distribution in $\{u_i : i\in \mathcal{J}\}$ for any $x$, the probability of the events $A_i\cap B_i$ will be independent of $x$ too. That means, for each pairwise disjunct event $A_i\cap B_i$ the probability will be same for any $y\in \mathbb{Z}_{2^{t-1}}^n$.
\\ So we could substitute the $B_i$-s with $\{ \sum_{j\in \mathcal{J}}u_j=2y\}$ for any $y\in \mathbb{Z}_{2^{t-1}}^n$, thus
$$P(\sum a_iu_i=2x)=P(\sum a_iu_i=2y)$$
concluding the proof.
\QEDB \\ \\
For the sake of completeness, we state a more general theorem: \\ \\
\textbf{Theorem 2.} \textit{Let $p$ be any prime, $k\in \mathbb{N}$, and $u_1,...,u_{n+1}$ a uniformly random sample from $\mathbb{Z}_{p^t}^n$. Let $A=\{ \textbf{a}=(a_1,...,a_{n+1})\ne (0,...0) \; :\; \sum_{i=1}^{n+1}a_iu_i\equiv 0 \; (mod \;p), a_i\in \{ -(p-1),...,-1,0,1,2,...,p-1\} \}$. Then, if we choose an $\textbf{a}\in A$ randomly, $\frac{\sum a_iu_i}{p}$ will be a uniformly random element from $\mathbb{Z}_{p^{k-1}}^n$.} \\ \\
\textbf{Proof} The proof is very similar to the previous case, only here we will have more $A_i\cap B_i$ events. For clarity, here $B_i$ is the event $\{ \sum_{j\in \mathcal{J}} a_ju_j= px\}$, for a given $x\in \mathbb{Z}_{p^{k-1}}^n$, and $\mathcal{J}$ consists of those $j$-s, for which $a_j\ne 0$ in the $i$-th element of $A$. The independence from $x$ will follow from the same argument and the fact that $a_iu_i\equiv px_i \; (mod \; p^2)$ will have a unique solution in $u_i$ for any $a_i\in \{ -(p-1),...,-1,+1,..,p-1\}$ and $x_i\in \mathbb{Z}_{p^{k-1}}$, because there exists a solution if and only if $gcd(a_i,p^2)|px_i$, which is trivially true, and the number of solutions has to be $gcd(a_i,p^2)$, which will always be $1$, because $p$ is prime. (So we can choose $|\mathcal{J}|-1$ $u_i$ freely and then the last one will be determined uniquely, so we have the same number of solutions for any $x$.) \QEDB
\\ \\
Next, we will give a method to get a uniformly random element from $A$:
\\
Observe, that $\sum a_iu_i\equiv 0$ $(mod$ $2)$ can be easily transformed to a set of linear equations over $\mathbb{Z}_2$, thus we can use Gaussian elimination to achieve echelon form. Once that is done, we will count the free variables. Let $k$ be the number we got. Then we choose a random element from $\{ 0,1\}^k\setminus 0^k$, and assign the free variables 0 or 1 accordingly. (If we would let all free variables to be 0, then the fixed ones had to be 0 too). Since any chosen element will correspond to a unique $\textbf{a}\in A$ and any $\textbf{a}\in A$ is achievable this way (because it has to be a solution), this method will give a uniform $\textbf{a}\in A$ indeed. \\ \\
Now we will clarify how we can move to a subgroup of $\mathbb{Z}_{2^t}^n$ once we know the parity of s.
\\ \\
\textbf{Lemma 2.} \textit{Let $G=\mathbb{Z}_{2^t}^n\rtimes \mathbb{Z}_2$ and let $f:G\to S$ be a function hiding the subgroup $H=\{ (0,0), (s,1)\}$. Then, if we know $s$ $(mod$ $2)$, we can reduce the problem to a subgroup $K\le G$ isomorphic to $\mathbb{Z}_{2^{t-1}}^n\rtimes \mathbb{Z}_2$.}
\\ \\
\textbf{Proof} Let $s_2$ denote $s$ $(mod$ $2)$. For each possible vector $s_2\in \mathbb{Z}_2^n$, let $K_{s_2}$ be the subgroup $\{ (2x,0) ,\; (y,1) \, : \; x\in \mathbb{Z}_{2^{t-1}}^n, \; y\equiv s_2 (mod \; 2)\}$, $y\equiv s_2 (mod \; 2)$ meaning that $y_i\equiv s_{2_i} (mod \; 2)$ for every $i=1,...,n$. \\
Observe that each $K_{s_2}$ has $2^{nt}$ elements, closed under multiplication and that the map $\phi :K_{s_2}\to \mathbb{Z}_{2^{t-1}}^n\rtimes \mathbb{Z}_2$, $\phi (2x,0)=(x,0)$, $\phi (s_2,1)=(0,1)$, $\phi (2x+s_2,1)=(x,1)$ is both a bijection and a homomorphism, so it is an isomorphism. \\
Furthermore, each $K_{s_2}$ contains every possible subgroup of the form $H=\{ (0,0), (s',1)\}$ with $s'\equiv s$ $(mod$ $2)$ and none with $s'\not\equiv s$, so we can safely and uniquely move to one of the $K_{s_2}$-s knowing $s_2$. \QEDB
\\ \\
\textbf{Remark.} An equivalent way to prove 
\textit{Lemma 2}
would be if we change $f:\mathbb{Z}_{2^t}^n\rtimes \mathbb{Z}_2\to S$  to $f':\mathbb{Z}_{2^{t-1}}^n\rtimes \mathbb{Z}_2\to S$ according to $s_2$, with $f'(x,0)=f(2x,0)$ and $f'(x,1)=f(2x+s_2,1)$. So now $f'(x,0)=f(2x,0)=f(2x+s,1)=f'(x+\frac{s-s_2}{2},1)$, hence $f'$ hides the subgroup $\{ (0,0), (\frac{s-s_2}{2},1)\}$. Now from $\frac{s-s_2}{2}$ $(mod$ $2)$ we can determine $s$ $(mod$ $4)$, and so forth.
\\ \\
For completeness, 
we describe Kuperberg's method\textsuperscript{\cite{DBLP:journals/siamcomp/Kuperberg05}} to obtain $|\phi_{u\pm u'}\rangle $ from $|\phi_u\rangle$ and $|\phi_{u'}\rangle$. First, we create the tensor product $|\phi_u\rangle \otimes |\phi_{u'}\rangle$ of the two states, which will be 
$$\frac{1}{2}(|00\rangle + e^{\frac{2\pi i\langle u,s\rangle}{N}}|10\rangle + e^{\frac{2\pi i\langle u',s\rangle}{N}}|01\rangle + e^{\frac{2\pi i\langle u+u',s\rangle}{N}}|11\rangle).$$
Then we apply a CNOT-gate and obtain 
$$\frac{1}{2}(|00\rangle + e^{\frac{2\pi i\langle u,s\rangle}{N}}|11\rangle + e^{\frac{2\pi i\langle u',s\rangle}{N}}|01\rangle + e^{\frac{2\pi i\langle u+u',s\rangle}{N}}|10\rangle).$$
Now we can measure the second qubit and get either $\frac{1}{\sqrt{2}}(|0\rangle +e^{\frac{2\pi i\langle u+u',s\rangle}{N}}|1\rangle)=|\phi_{u+u'}\rangle$ or $\frac{1}{\sqrt{2}}(e^{\frac{2\pi i\langle u,s\rangle}{N}}|1\rangle + e^{\frac{2\pi i\langle u',s\rangle}{N}}|0\rangle )=\frac{1}{\sqrt{2}}(|0\rangle +e^{\frac{2\pi i\langle u-u',s\rangle}{N}}|1\rangle )=|\phi_{u-u'}\rangle$ with equal probability.
\subsection{Complexity of our algorithm}
Now we finish the proof of Theorem 1 by analysing the time and space required for our algorithm. 
\\
Step 1 requires time  $\mathcal{O}(t^3(n+1)^4l)$ (because the QFT over $\mathbb{Z}_{2^t}^n$ has complexity $\mathcal{O}((nt)^2)$, and has to be done to prepare each $|\phi_u\rangle $) and space $\mathcal{O}(tn^2+l)$ (since we can use the same $l$ qubits for generating each $(u,|\varphi_u \rangle )$ pair) to prepare the  $\mathcal{O}(n)$ states needed for the algorithm. As we showed before most of this can be implemented on a classical computer, so the needed quantum space is only $\mathcal{O}(nt+l)$. Finding the coefficients $a_i$-s with the above written technique requires $\mathcal{O}((n+1)^3)$ time for each $(n+1)$-tuple, and during step 1 to step 5 we will apply it $(n+t)(n+1)^{t-2}$+$(n+t)(n+1)^{t-3}$+...+$(n+t)$ times, which combined can be computed in $\mathcal{O}(t(n+1)^{t+2})$ time. Then, obtaining the $(n+t)$ $|\phi_{v^{(t-1)}} \rangle$-s can be done with $\mathcal{O}((n+t)(n+1)^{t-1})$ additions, each requiring $\mathcal{O}(1) $ time (a CNOT-gate and a measurement).
Step 4, if we have $n$ linearly independent equations, can be solved in $\mathcal{O}((n+t)^3)$ time too with Gaussian elimination.
\\ 
Thus, the algorithm computes $s$ $(mod$ $2)$ in $\mathcal{O}(t^3(n+1)^4l)+\mathcal{O}(t(n+1)^{t+2})\le \mathcal{O}(t^3(n+1)^{t+2}l)$ time, and then iterating it for $\mathbb{Z}_{2^{t-1}}^n$,...,$\mathbb{Z}_2^n$ takes $\mathcal{O}(t^3(n+1)^{t+1}l)$,...,$\mathcal{O}(t^3(n+1)^{3}l)$ time respectively. So the whole algorithm has running time  $\mathcal{O}(t^3(n+1)^{t+2}l)$ ( or $\mathcal{O}(t^3(n+1)^{(t+2)}l)\log (\varepsilon^{-1} ),$ if we need $\varepsilon$ error instead of $\frac{1}{2}$) and requires $\mathcal{O}(tn^2+l)$ classical and $\mathcal{O}(nt+l)$ quantum space (step 1 to step 5 can be computed in $\mathcal{O}(tn^2+l)$ classical and $\mathcal{O}(nt+l)$ quantum space and we can use the same space for each iteration as we only need $\mathcal{O}(nt) $ space to store the actual values of $s$, which can be overwritten after each iteration too).
\subsection{A modified version of our algorithm with coset-state input}
Finally, another advantage of our algorithm compared to the FIMSS is that a slightly modified version can be implemented with coset-states as input, which are (in our case) states of the form $\frac{1}{\sqrt{2}}(|x\rangle |0\rangle +|x+s\rangle |1\rangle )$. Its usefulness is that it can use these states, even if they are generated by an other source. 
\\ The modified version of our algorithm is as follows:
\\In the Input we suppose that we have a stream of coset states. We could have all the required states as input, but that would need an exponential amount of space. Then we apply the QFT to these states and follow the steps of the original algorithm. Only the iteration after obtaining $s \; (mod \;2^i)$ will be different: \\
After we know $s_i=s \; (mod\; 2^i)$ we will start with applying a unitary transform $U$ to the coset states, which maps $|x\rangle |i\rangle \to |x-i*s_i\rangle |i\rangle$, $i=0,1$, and only then use the QFT ($U$ is unitary, since it permutes the basis vectors $|x\rangle |i\rangle $, $x\in \mathbb{Z}_{2^t}^n$, $i\in \{ 0,1\}$). 
\\ So now our $|\phi_v\rangle$ states will have the form $$\frac{1}{\sqrt{2}}(|0\rangle +e^{\frac{2\pi i \langle v,s-s_i\rangle}{N}}|1\rangle )$$
Since $s-s_i$ is divisible by $2^i$, we only need to repeat the first three steps until we have $n+t$ $|\phi_{v^{(t-i-1)}}\rangle$ states, where $v^{(t-i-1)}$'s coordinates are divisible by $2^{t-i-1}$, because now the scalar product will be a multiple of $2^{t-1}$. That means, if we measure the state in the $\pm$ basis, we will obtain equations for the parity of the sum of some coordinates of $\frac{s-s_i}{2^i}$ (the coordinates being the ones for which the same coordinate in $u^{(t-i-1)}=\frac{v^{(t-i-1)}}{2^{t-i-1}}$ is odd), so we can determine $\frac{s-s_i}{2^i} \; (mod \; 2)$ the same way and acquire $s_{i+1}$.
\section{A tailored version of the FIMSS for  $\mathbb{Z}_{2^t}^n$}

Although running the
FIMSS algorithm for the hidden shift problem in groups of the form  
$\mathbb{Z}_{p^t}^n$ would require at least 
$\mathcal{O}(n^{pt})poly(n,t,l)$ time and space, in the special case of  
$\mathbb{Z}_{2^t}^n$ with appropriate modifications the running time 
can be lowered to $(\mathcal{O}(n))^{t+2}poly(n,t,l)$, 
so it will be similar to our algorithm's.
This part is devoted to a more or less self-contained outline of
such a modified version. 
Note however that the space requirements remain exponential in $t$, 
so the method of this paper could be a lot easier to implement, 
therefore it should still be a useful recursive tool for the general case.

We start with recalling some basic concepts from 
Ref. \cite{DBLP:journals/siamcomp/FriedlIMSS14} with appropriate
simplifications.
A simple but important idea is formulating a version of the hidden 
shift problem that is suitable for taking "averages" over
subgroups and making recursions into factor
groups possible. Let $\Gamma$ be a finite set of mutually orthogonal quantum states and let
$G$ be a finite group.
A map $\alpha : G\times \Gamma \to \Gamma$ is said to be a 
\textit{group action} if for every $x\in G$ the map $\alpha_x : 
|\phi\rangle \to |\alpha (x,|\phi \rangle )\rangle$ gives a 
permutation of $\Gamma$, $\alpha_{1_G}$ is the identity map and 
$\alpha_x \circ \alpha_{y^{-1}}=\alpha_{xy^{-1}}$. 
We denote $|\alpha (x,|\phi \rangle )\rangle$ 
by $|x\cdot \phi \rangle$ and $\sum_{x\in G}|x\cdot \phi \rangle$ by $|G\cdot \phi\rangle$. \\
We assume that the action $\alpha$ is given by an oracle.
The {\em single translator} problem assumes that $G$ acts (semi-)regularly on $\Gamma$
and "sufficiently many" copies of a pair 
$|\phi_0 \rangle,|\phi_1\rangle$ from
$\Gamma$ as input. 
The output should be the element $x\in G$ 
such that $|\phi_0 \rangle=|x\phi_1\rangle$ (provided existence).
(This problem is a simplification of the more general {\em translating coset}
problem of Ref. \cite{DBLP:journals/siamcomp/FriedlIMSS14} which
is for the situation when the action is not semiregular.)
To 
capture the hidden shift problem for a function
$f:G\rightarrow S$ consider the superpostion 
$|f\rangle=\sum_{z\in G}|z\rangle|f(z)\rangle$. Then
the corresponding superposition $|xf\rangle$ for the shifted function $xf:z\mapsto f(zx)$
can be obtained by multiplying the first register by $x^{-1}$
from the right:
$\sum_{z\in G}|z\rangle|f(zx)\rangle =
\sum_{z\in G}|zx^{-1}\rangle |f(z)\rangle$.

As already mentioned, the input for the {\em single translator} problem
consists of several identical copies of the pair of states.
Formally, 
we have $|\phi_0\rangle^{\otimes k} \otimes |\phi_1\rangle^{\otimes k}$
as input for some integer $k$ (which needs to be
sufficiently large to get a correct answer with reasonable
probability). Our averages will be superpositions of
orbits on several identical copies
of states. Formally, let 
$\Gamma^k=\{ |\phi\rangle^{\otimes k}: |\phi\rangle \in \Gamma \}$,
the set of products of $k$ identical copies of states from $\Gamma$
and define $\alpha^k$ on 
$\Gamma^k=\{ |\phi\rangle^{\otimes k}: |\phi\rangle \in \Gamma \}$ 
as $\alpha^k(x,|\phi \rangle^{\otimes k})=|x\cdot \phi \rangle^{\otimes k}$. 
Obviously, if we have an oracle for 
$\alpha$ (that maps 
$|x\rangle |\phi \rangle \to |x\rangle |x\cdot \phi \rangle$ ) 
then we can simulate an oracle for $\alpha^k$ 
too using $k$ queries to the oracle for $\alpha$.

Let us turn to the special case $G=\mathbb{Z}_{2^t}^n$. Then we can take a 
length $t$ 
subnormal series 
of $G$, where each factor group will be isomorphic to $\mathbb{Z}_2^n$ and in these groups the subroutine for
\textit{single translator} reduces essentially to Simon's problem.
\\
The algorithm will mostly be the same as the standard FIMSS, so the reader is invited to read Ref. 
\cite{DBLP:journals/siamcomp/FriedlIMSS14} for details 
omitted here and for further clarification.
Let $N=2^{t-1}G\cong
\mathbb{Z}_2^n$ and
$|F_i\rangle = |\phi_i\rangle^{\otimes q}$, where $q$ is some integer we set
in advance. 
\\
The \textit{orbit} of a state $|\phi \rangle$ under a subgroup $K\leq G$ 
is 
$K(|\phi \rangle )=\{ |x\cdot \phi \rangle : x\in K\}$,
while the \textit{orbit superposition} is the uniform superposition
$|K\phi \rangle =\sum_{x\in K}|x\cdot \phi\rangle$.

We can modify 
the exact algorithm of Brassard and Hoyer\textsuperscript{\cite{595153}} for  
Simon's problem to take $q=O(n)$ identical copies of 
two quantum states as input in the same way as in Ref. 
\cite{DBLP:journals/siamcomp/FriedlIMSS14} 
Corollary 3.6 and 3.8.
From that we can construct an algorithm for the \textit{single translator} problem
 in $N\cong \mathbb{Z}_2^n$, that maps $|x\cdot \phi\rangle^{\otimes
q}|\phi\rangle^{\otimes q}|0\rangle $ to 
$|x\cdot \phi\rangle^{\otimes q}|\phi\rangle^{\otimes q}|x\rangle$ 
like the ElementaryAbelianTCS algorithm in Ref. \cite{DBLP:journals/siamcomp/FriedlIMSS14}.
\\ \\
Based on this, we give an algorithm for creating the uniform superposition 
of a given state's orbit under some 
group action $\alpha$ of $N$ 
(so creating the state $|N\cdot \phi \rangle$). Call it
\textbf{OS}$(N,\alpha,q,\phi)$.
\\ It takes as input $|\phi\rangle^{\otimes 2q}|N\rangle$.
\\
\textbf{Step 1:} Apply the group element in the last register to the first $q$ registers:
$$\sum_{x\in N}|x\cdot \phi \rangle^{\otimes q}|\phi \rangle^{\otimes q}|x\rangle$$
\textbf{Step 2:} Apply the translator finding algorithm (the one that maps $|x\cdot \phi\rangle^{\otimes
q}|\phi\rangle^{\otimes q}|0\rangle $ to 
$|x\cdot \phi\rangle^{\otimes q}|\phi\rangle^{\otimes q}|x\rangle$ ) for $\mathbb{Z}_2^n$ backwards:
$$|NF\rangle |F\rangle |0\rangle,$$
where $|F\rangle=|\phi\rangle^{\otimes q}$
and $|NF\rangle$ is the orbit superposition under the action
$\alpha^q$ on $N$.
\\
Now let's get to the main algorithm: \textbf{T}$(G,N,\alpha,q(r+1))$.\\
If $t=1$, so $G=\mathbb{Z}_2^n$ then we saw that we can solve the 
problem in time $poly(n)$ for $q=O(n)$.
So let us suppose by induction, that we already have an 
algorithm for \textbf{T}$(G/N,\alpha,r)$ ($G/N\cong\mathbb{Z}_{2^{t-1}}^n$ 
and we present how to solve it in $G=\mathbb{Z}_{2^t}^n$ 
using that as a subroutine).
The algorithm takes as input 
$|\phi_0\rangle^{\otimes q(r+1)}|\phi_1\rangle^{\otimes q(r+1)}|0\rangle$ 
with ancilla $|N\rangle^{\otimes 2r}|\textbf{0}\rangle$.\\
\\ \textbf{Step 1:} We call the algorithm \textbf{OS}$(N,\alpha ,q,\phi_i)$ $r$ times on blocks of the form
$|\phi_0\rangle^{\otimes q}|N\rangle$ and $r$ times on blocks $|\phi_1\rangle^{\otimes q}|N\rangle$ to get: 
$$|N\cdot F_0\rangle^{\otimes r}|F_0\rangle |N\cdot F_1\rangle^{\otimes r}|F_1\rangle
|0\rangle |0\rangle^{\otimes 2r}|\textbf{0}\rangle .$$
\textbf{Step 2:} We recursively call \textbf{T}$(G/N,N,\alpha^q,r)$ on $|N\cdot F_0\rangle^{\otimes r}|N\cdot F_1\rangle^{\otimes
r}|0\rangle$ (We can do it, since $G/N\cong \mathbb{Z}_{2^{t-1}}^n$ and we already have our algorithm for that for any $\alpha$ group action):
$$|N\cdot F_0\rangle^{\otimes r}|F_0\rangle |N\cdot F_1\rangle^{\otimes r}|F_1\rangle
|0\rangle |0\rangle^{\otimes 2r}|sN\rangle.$$
\textbf{Step 3:} Undo Step 1:
$$|\phi_0\rangle^{\otimes q(r+1)}|\phi_1\rangle^{\otimes q(r+1)}|0\rangle |N\rangle^{\otimes 2r}|sN\rangle.$$
Now, since $|s\cdot \phi_0\rangle =|\phi_1\rangle \Longleftrightarrow s=vu$, $u\in N$, such that $|u\cdot
\phi_0\rangle = |v^{-1}\cdot \phi_1\rangle$, we reduced the problem to a simple hidden shift problem in $\mathbb{Z}_2^n$, so with one more use of the modified 
Brassard--Hoyer
algorithm we can find $u$, then $s$.
\\
Therefore, because the Brassard-Hoyer algorithm for Simon's problem 
only needs $q=\mathcal{O}(n)$ states to work, then by simple induction we can conclude, that the algorithm needs $(\mathcal{O}(n))^t$ states, because one level would need only $\mathcal{O}(n)$, therefore by induction we have to give $\mathcal{O}(n)$ blocks of $(\mathcal{O}(n)^{t-1})$ states to the recursive calls.
Then taking in factor that the final Brassard-Hoyer algorithm needs $O(n)$ states too at the end, and that the 
$|\phi_i\rangle=\sum_x|x\rangle|f_i(x)\rangle$ states each take $O(nt)$ qubits, we have that both the space and time complexity of the algorithm will be $poly(n,t,l)(\mathcal{O}(n))^{t+2}$, so it should be similar, if not somewhat slower than our method. The subroutines that the algorithm uses, like the \textbf{OS} can all be implemented in polynomial time, and can be done simultaneously on the blocks, so they don't really affect the running time.

\section*{Acknowledgements}
The author
would like to thank Gábor Ivanyos for 
helpful remarks and suggestions.
\printbibliography

@article{DBLP:journals/aam/EttingerH00,
  author    = {Mark Ettinger and
               Peter H{\o}yer},
  title     = {On Quantum Algorithms for Noncommutative Hidden Subgroups},
  journal   = {Adv. Appl. Math.},
  volume    = {25},
  number    = {3},
  pages     = {239--251},
  year      = {2000},
  url       = {https://doi.org/10.1006/aama.2000.0699},
  doi       = {10.1006/aama.2000.0699},
  bibsource = {dblp computer science bibliography, https://dblp.org}
}

@article{DBLP:journals/siamcomp/FriedlIMSS14,
  author    = {Katalin Friedl and
               G{\'{a}}bor Ivanyos and
               Fr{\'{e}}d{\'{e}}ric Magniez and
               Miklos Santha and
               Pranab Sen},
  title     = {Hidden Translation and Translating Coset in Quantum Computing},
  journal   = {{SIAM} J. Comput.},
  volume    = {43},
  number    = {1},
  pages     = {1--24},
  year      = {2014},
  url       = {https://doi.org/10.1137/130907203},
  doi       = {10.1137/130907203},
  bibsource = {dblp computer science bibliography, https://dblp.org}
}

@article{DBLP:journals/siamcomp/Kuperberg05,
  author    = {Greg Kuperberg},
  title     = {A Subexponential-Time Quantum Algorithm for the Dihedral Hidden Subgroup
               Problem},
  journal   = {{SIAM} J. Comput.},
  volume    = {35},
  number    = {1},
  pages     = {170--188},
  year      = {2005},
  url       = {https://doi.org/10.1137/S0097539703436345},
  doi       = {10.1137/S0097539703436345},
  bibsource = {dblp computer science bibliography, https://dblp.org}
}

@article{DBLP:journals/siamcomp/Shor97,
  author    = {Peter W. Shor},
  title     = {Polynomial-Time Algorithms for Prime Factorization and Discrete Logarithms
               on a Quantum Computer},
  journal   = {{SIAM} J. Comput.},
  volume    = {26},
  number    = {5},
  pages     = {1484--1509},
  year      = {1997},
  url       = {https://doi.org/10.1137/S0097539795293172},
  doi       = {10.1137/S0097539795293172},
  bibsource = {dblp computer science bibliography, https://dblp.org}
}

@article{article,
author = {Regev, Oded},
year = {2004},
month = {07},
pages = {},
title = {A Subexponential Time Algorithm for the Dihedral Hidden Subgroup Problem with Polynomial Space}
}

@article{doi:10.1137/S0097539703440678,
author = {Regev, Oded},
title = {Quantum Computation and Lattice Problems},
journal = {SIAM Journal on Computing},
volume = {33},
number = {3},
pages = {738-760},
year = {2004},
doi = {10.1137/S0097539703440678},

URL = { 
        https://doi.org/10.1137/S0097539703440678
    
},
eprint = { 
        https://doi.org/10.1137/S0097539703440678
    
}

}

@INPROCEEDINGS{595153,
  author={G. {Brassard} and P. {Hoyer}},
  booktitle={Proceedings of the Fifth Israeli Symposium on Theory of Computing and Systems}, 
  title={An exact quantum polynomial-time algorithm for Simon's problem}, 
  year={1997},
  volume={},
  number={},
  pages={12-23},
  doi={10.1109/ISTCS.1997.595153},
  ISSN={},
  month={June},}
\end{document}